\documentclass[journal]{IEEEtran}
\usepackage{graphicx}
\usepackage{amsfonts}
\usepackage{amsmath}
\usepackage{cite}
\usepackage{multirow}
\usepackage{url}
\usepackage{colortbl}

\ifCLASSINFOpdf
\else
\fi
\begin{document}
%
\title{L3C-Stereo: Lossless Compression for Stereo Images}
%
%
%
\author{Zihao~Huang,
        Zhe~Sun,
        Feng~Duan,~\IEEEmembership{Member,~IEEE,}
        Andrzej~ Cichocki,~\IEEEmembership{Fellow,~IEEE, }
        Peiying~Ruan,
        and Chao~Li
\thanks{Z. Huang and F. Duan are with the College of Artificial Intelligence, Nankai University, Tianjin, China.
Z. Sun is with the Computational Engineering Applications Unit, Head Office for Information Systems and Cybersecurity, RIKEN, Saitama, Japan.
A. Cichocki is with the Skolkovo Institute of Science and Technology, Moscow, Russia; College of Computer Science, Hangzhou Dianzi University, China; Department of Informatics, Nicolaus Copernicus University, Poland and the Systems Research Institute of Polish of Academy of Science, Warsaw, Poland.
P. Ruan is with the NVIDIA AI Technology Center, NVIDIA Corporation Japan, Japan.
C. Li is with the Center for Advanced Intelligence Project (AIP), RIKEN, Japan.
}
\thanks{Zihao Huang and Zhe Sun contributed equally to this work}
\thanks{Corresponding author: Feng Duan (duanf@nankai.edu.cn)}
}

%
%

\markboth{Journal of \LaTeX\ Class Files,~Vol.~14, No.~8, August~2015}%
{Shell \MakeLowercase{\textit{et al.}}: L3C-Stereo: Lossless Compression for Stereo Images}
%



\maketitle

\begin{abstract}
A large number of autonomous driving tasks need high-definition stereo images, which requires a large amount of storage space. Efficiently executing lossless compression has become a practical problem. Commonly, it is hard to make accurate probability estimates for each pixel. To tackle this, we propose L3C-Stereo, a multi-scale lossless compression model consisting of two main modules: the warping module and the probability estimation module. The warping module takes advantage of two view feature maps from the same domain to generate a disparity map, which is used to reconstruct the right view so as to improve the confidence of the probability estimate of the right view.
The probability estimation module provides pixel-wise logistic mixture distributions for adaptive arithmetic coding. In the experiments, our method outperforms the hand-crafted compression methods and the learning-based method on all three datasets used. Then, we show that a better maximum disparity can lead to a better compression effect. Furthermore, thanks to a compression property of our model, it naturally generates a disparity map of an acceptable quality for the subsequent stereo tasks.
\end{abstract}


%
\IEEEpeerreviewmaketitle

\section{Introduction}
With the improvement of autonomous driving technologies and the proposal of superior algorithms, higher quality stereo images are required, which means that more storage space is needed. Commonly, it requires much more space to store a lossless image than a lossy one\cite{andriani2004comparison}. As a consequence, knowing how to compress stereo images in a better lossless way has become both essential and economically important.

Lossless compression has a long history of development. Using this method, it is possible to reconstruct the stored bitstream into an image that looks exactly like the original. Although non-learning-based methods such as PNG\cite{png}, WebP\cite{webp} and FLIF\cite{sneyers2016flif} are widely used, learning-based methods can achieve higher a compression ratio. Methods \cite{mentzer2019practical,cao2020lossless,mentzer2020learning,hoogeboom2019integer,van2020idf++,zhang2020lossless} all consist of a probability estimation model and an entropy codec. In the encoding phase, the probability of all the symbols appearing at each subpixel is modelled, followed by the images being encoded to a bitstream by the entropy codec. The key to lossless compression is the accuracy of the probability estimation; the more accurate it is, the higher the compression ratio achieved is.

Stereo images contain both left and right views captured at the same time, where similar information naturally exists. Many studies \cite{kendall2017end,chang2018pyramid,yang2018segstereo,duggal2019deeppruner,guo2019group} have leveraged this information estimate disparity map \footnote{A disparity map refers to the apparent pixel difference or motion between a pair of stereo images.}for downstream tasks. This similar information was first used for deep lossy compression in \cite{liu2019dsic}. By estimating a disparity map,  left view feature maps were warped to the right ones to achieve a better lossy compression. However, as far as we know, there is no existing research on lossless compression for stereo images.

In this paper, we address this challenge and propose a learning-based architecture for lossless stereo image compression, as illustrated in Fig.~\ref{L3C-stereo}. In order to extract useful information from left to right view, a relatively accurate estimation of the disparity map is needed, since the values in a disparity map represent the pixel distances in the two view images. Many studies have shown that hierarchical structures can improve the disparity map estimation. As a consequence, we adopt the L3C \cite{mentzer2019practical}, a hierarchically learned lossless compression structure, as our backbone to then estimate the disparity map under multi-scaled stereo feature maps to achieve a better right view compression. 

Specifically, given a pair of stereo images $\boldsymbol{X}_L$ and $\boldsymbol{X}_R$, we employ $S$ encoders $\texttt{E}^{(s)}$ for feature extraction, yielding the multi-scale auxiliary features $\boldsymbol{Z}_L^{(s)}$ and $\boldsymbol{Z}_R^{(s)}$ which will be encoded to a bitstream and transferred to the receiver. Then, $S$ decoders $\texttt{D}^{(s)}$ are employed to integrate current scale features with lower scale features. For the left view, the decoded feature maps are used to predict the probability distributions $p$ for $\boldsymbol{X}_L$ and $\boldsymbol{Z}_L^{(s)}$. For the right view, the disparity map is leveraged to flow information from the left view to the right view. After that, the fine-tuned decoded feature maps allow us to get a more accurate probability distribution for the right view. In the end, the sender uses a codec to encode all the images and auxiliary features to a bitstream. Regarding the receiver, the codec needs the same probability distributions $p$ to decode the bitstream. Consequently, it is necessary to reconstruct the distributions at the receiver. Starting with the smallest scale auxiliary feature $\boldsymbol{Z}_L^{(S)}$ and $\boldsymbol{Z}_R^{(S)}$, we assume the prior distribution to be uniform both in the sender and the receiver. This allows the codec to decode $\boldsymbol{Z}_L^{(S)}$ and $\boldsymbol{Z}_R^{(S)}$ in a non-optimal way, with a low cost due to its low scale. Finally, all the images and features can be decoded from a low to a high scale.

The proposed algorithm is evaluated on the KITTI 2012/2015 and Scene Flow datasets. Experiments show that our method outperforms the baseline and other lossless image compression methods. Furthermore, the effects on compression by the supervised/self-supervised disparity map estimation and by the maximum disparity are also investigated. Finally, note that the disparity map which made as a byproduct of our method can be easily used for the follow-up automatic driving tasks.

\section{Related Works}

\subsection{Learned lossy compression}
Lossy compression \cite{hu2021learning} has a higher compression ratio, but the resulting image quality is poorer. Given an image $\boldsymbol{x}$, an encoder and a quantizer are used to generate a discrete representation $\boldsymbol{y}$, which will be encoded to a bitstream for transfer. Then, a decoder will be used to generate an output $\widehat{\boldsymbol{x}}$ which is similar to $\boldsymbol{x}$. In the training phase, it is hoped that this model can achieve a high compression ratio while being able to produce high quality images. Considering the trade-off, the loss function is usually of this form:
\begin{equation}
L=l(\boldsymbol{x},\widehat{\boldsymbol{x}})+\beta R(\boldsymbol{y}).
\end{equation}
Where $l(\cdot )$ is the reconstruction error and $R(\boldsymbol{y})$ is the cost of encoding $\boldsymbol{y}$ to a bitstream. Ball\'{e} \textit{et al.}\cite{balle2016end,balle2018variational,minnen2018joint} proposed the generalized divisive normalization (GDN) and several CNN-based variational autoencoders for learned lossy compression, introducing the side information to obtain a better reconstruction. Chen \textit{et al.}\cite{chen2021end} used self-attention to improve the compression effect. References \cite{johnston2018improved,toderici2017full,lin2020spatial} implemented lossy compression using RNN. Liu \textit{et al.}\cite{liu2019dsic} also made use of the side information for stereo image lossy compression, but the information was from the left view image rather than from the extracted representation.

\subsection{Learnable lossless compression}
Lossless compression does not usually have a high compression ratio, because high frequency components or noise in the image are difficult to compress. All the proposed methods need to estimate a distribution on the original image $\boldsymbol{x}$. The pixel-wise auto-regressive model explored in \cite{oord2016conditional,van2016pixel,salimans2017pixelcnn++} predicts the distribution of each pixel in turn conditionally:
\begin{equation}
p\left ( \boldsymbol{x} \right )=\prod_{i}p\left ( \boldsymbol{x}_i|\boldsymbol{x}_1,\boldsymbol{x}_2,...,\boldsymbol{x}_{i-1} \right ),
\end{equation}
where $i$ is the pixel index. The multi-scale auto-regressive model \cite{cao2020lossless,mentzer2019practical,zhang2020lossless,mentzer2021neural} estimates the distribution of pixels jointly based on low-scale auxiliary representations:
\begin{equation}
p\left ( \boldsymbol{x}_0,\boldsymbol{x}_1,\boldsymbol{x}_2,...,\boldsymbol{x}_N \right )=\prod_{n=0}^{N-1}p\left ( \boldsymbol{x}_n|\boldsymbol{x}_{n+1},\boldsymbol{x}_{n+2},...,\boldsymbol{x}_N \right )\cdot p\left ( \boldsymbol{x}_N \right ),
\end{equation}
where $\boldsymbol{x}_0$ is the original image while the other $\boldsymbol{x}_n$ are the low-scale auxiliary representations. Such methods are generally much faster than the previous methods, benefiting from a joint estimation rather than a conditional estimation.  Flow-based models\cite{hoogeboom2019integer,van2020idf++} learn an invertible mapping function $f(\cdot)$ from image space $\mathbb{X}$ to latent space $\mathbb{Z}$, where the variables follow a predefined distribution $p_{\boldsymbol{z}}$:
\begin{equation}
p\left ( \boldsymbol{x} \right )=p_{\boldsymbol{z}}\left ( f\left ( \boldsymbol{x} \right ) \right )\left | det\left ( \frac{\partial f\left ( \boldsymbol{x} \right )}{\partial \boldsymbol{x}} \right ) \right |.
\end{equation}
However, this kind of method is more suitable for low-resolution images. Moreover, some studies \cite{mentzer2020learning,bai2021learning} have employed lossy compression and residual lossless compression while achieving competitive results.

\subsection{Stereo matching}
A variety of studies have been done on stereo matching based on deep learning, and all of them can be roughly grouped into two types of network structures and two types of training methods. For the structure type, one \cite{mayer2016large,ilg2018occlusions,liang2018learning,pang2017cascade} uses the 2D-CNN-based encoder-decoder to compute the correlation scores between the left and right view features, while the other\cite{kendall2017end,chang2018pyramid,yang2018segstereo,duggal2019deeppruner,guo2019group} uses 3D-CNN to estimate the disparity from a 4D cost volume. As for the training methods, most studies have chosen to use supervised learning, given a ground-truth disparity map as a label. However, the ground-truth disparity is not always available. Hence, the other method \cite{zhong2017self,jie2018left} is self-supervised. Here, the network needs to warp the left view to the right view based on the estimated disparity map while minimising the reconstruction error. The disadvantage of this method is that different homologous pixel values and occlusion areas will lead to larger estimation errors. We will show the effect of the model under different supervised methods.

\section{Methods}

\begin{figure*}[ht] 
    \centering 
    \includegraphics[width=0.9\textwidth]{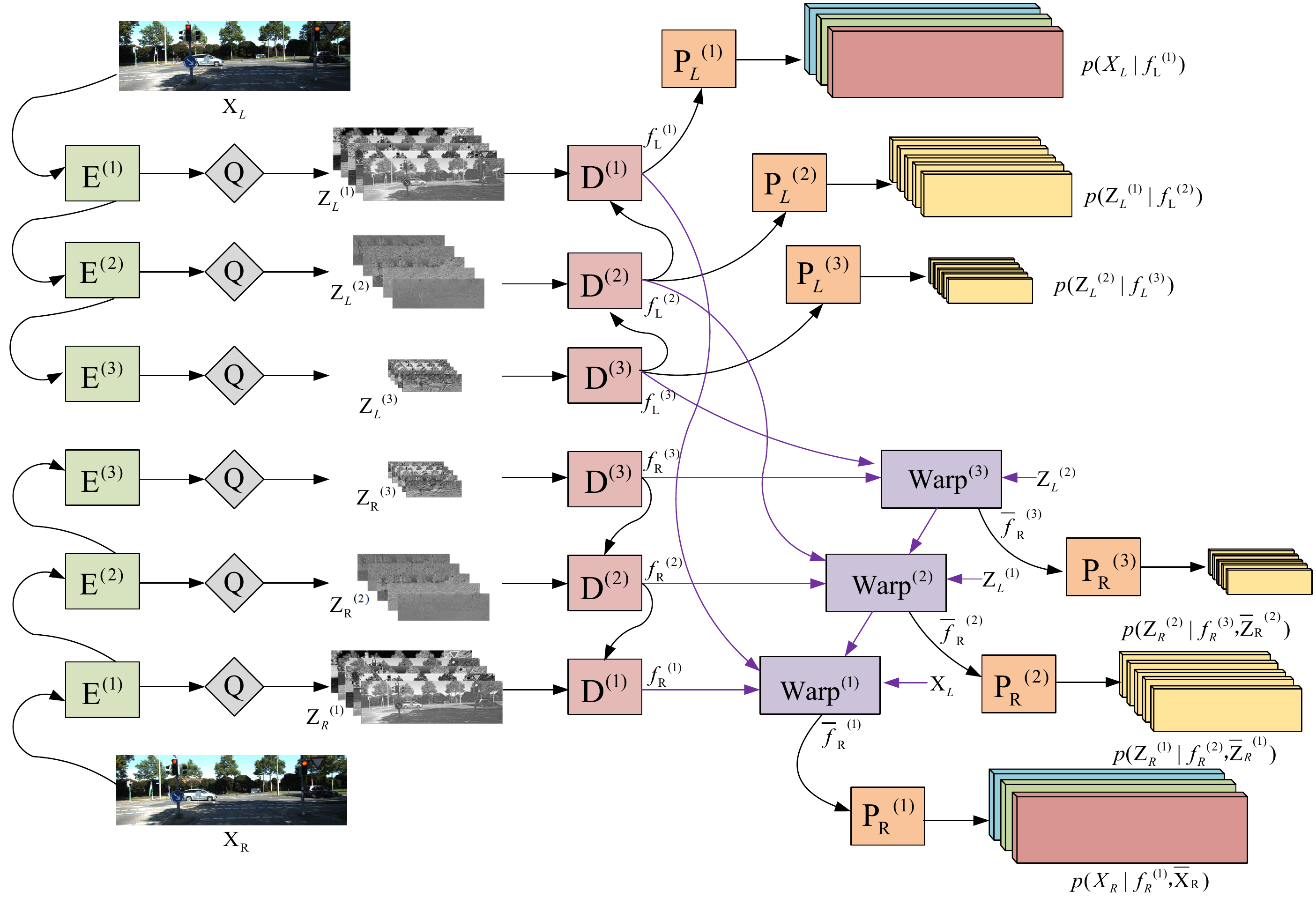} 
    \caption{An overview of the L3C-Stereo architecture. $\texttt{E}^{(s)}$ and $\texttt{D}^{(s)}$ are the encoder and decoder modules, respectively. The left and right view images $\boldsymbol{X}_L$ and $\boldsymbol{X}_R$ share the same $\texttt{E}^{(s)}$ and $\texttt{D}^{(s)}$. $\texttt{Q}$ is the quantization module, which is a non-learning module. The $\boldsymbol{Z}^{s}$s are the auxiliary feature maps in different scales. Module $\texttt{Warp}^{s}$ receives the $\boldsymbol{f}^{(s)}$ from the left and right view feature maps and the disparity map from $\texttt{Warp}^{(s+1)}$. It then estimates a disparity map for the right view, and finally it warps $\boldsymbol{Z}_{L}^{(s-1)}$ into $\overline{\boldsymbol{Z}}_{R}^{(s-1)}$, which is similar to $\boldsymbol{Z}_{R}^{(s-1)}$. $\texttt{P}^{(s)}$ is the probability estimation module.} 
    \label{L3C-stereo} 
\end{figure*}

\subsection{Lossless compression}
Commonly, given a stream of symbols $x_{1},x_{2}...x_{N}$, the value of each symbol $x_{i}$ is obtained from a set $\mathcal{S} =\left \{ 0, 1, 2, ..., S \right \}$ ($S=255$ for image symbol) based on a probability mass function $\widetilde{p}$. Knowing this function $\widetilde{p}$, we can encode the stream into a bitstream, and the entropy of $\widetilde{p}$, which equals the number of bits required to encode $x_{i}$, is
\begin{equation}
H\left ( \widetilde{p} \right )=\mathbb{E}_{x_{i}\sim \widetilde{p}}\left [ -log_{2}\widetilde{p}\left ( x_{i} \right ) \right ].
\end{equation}
However, $\widetilde{p}$ is not known, and we can only estimate a $p$ function that is close to $\widetilde{p}$, achieving a bitcost of:
\begin{equation}
H\left ( \widetilde{p},p \right )=\mathbb{E}_{x_{i}\sim \widetilde{p}}\left [ -log_{2}p\left ( x_{i} \right ) \right ].
\label{cross-entropy}
\end{equation}
As a simple example, we can assume that we have a pixel $x$ with a value of 20. Since we know exactly what the value of the pixel is, as a result $\widetilde{p}_{x}(x=20)=1$ and $\widetilde{p}_{x}(x\neq 20)=0$. With this prior knowledge, in fact no bits are required to encode $x$. If we get a $p_{x}$ obeying a uniform distribution, then $H\left ( \widetilde{p}_{x},p_{x} \right )=-log_{2}\left ( 1/255 \right )=7.99$, which means that using $p_{x}$ to encode $x$ the average bitcost is 7.99. Consequently, the more accurate the probability mass function used is, the higher the compression ratio of the model will be. 

\textbf{Arithmetic Coding (AC)} is one of the main algorithms used in lossless image compression, and it is a form of entropy encoding. In simple terms, based on a given probability mass function, AC divides the interval $[0,1)$ into multiple subintervals. Each subinterval represents a symbol, and the size of the subinterval is proportional to the probability of this symbol appearing. In the beginning, $L$ is set to 0 and $H$ is set to 1. Then, the symbols in the stream are read in turn, updating the upper and lower bounds of the subinterval $[L,H)$. Finally, any decimal in the obtained subinterval is outputted in binary form to obtain the bitstream. In this study, Adaptive Arithmetic Coding (AAC) was used to encode the image stream, since there is a high correlation among nearby pixels and thus an adaptive form can achieve a better compression.

\begin{figure*}[ht] 
    \centering 
    \includegraphics[width=1\textwidth]{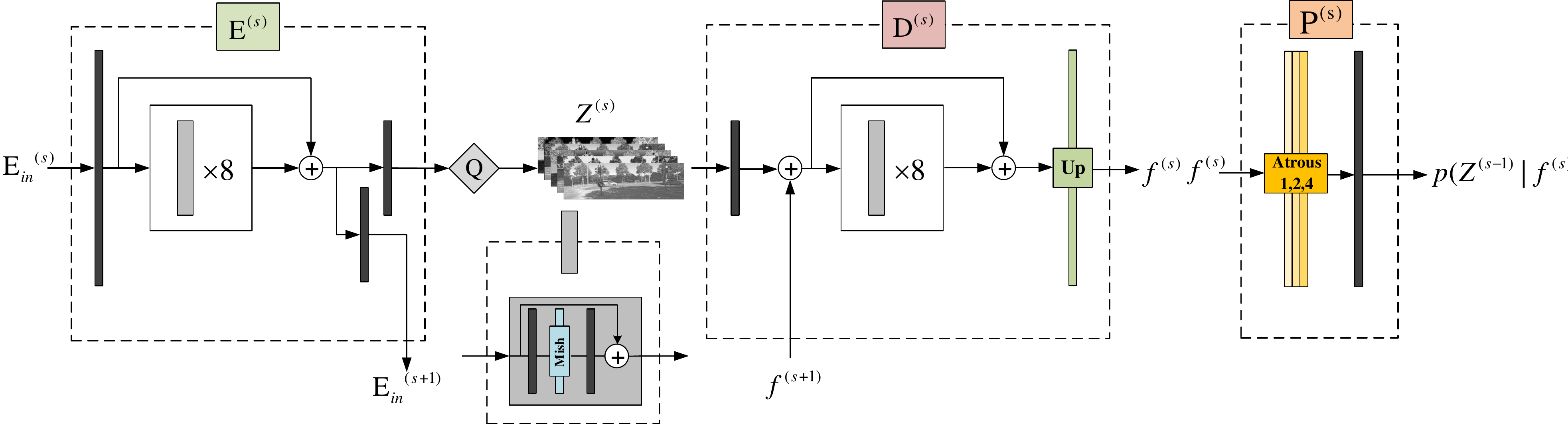} 
    \caption{Architecture details for $\texttt{E}^{(s)}$, $\texttt{D}^{(s)}$ and $\texttt{P}^{(s)}$. The contents within the dashed rectangles are the details of the respective modules or blocks indicated above them. All the black thin rectangles represent a 2D Conv. layer. Their heights represent the input size and the output channels are $64$ except for the one before module $\texttt{Q}$. The green thin rectangle is a pixel shuffling upsampling. The three yellow thin rectangles are 2D Conv. layers with different dilation rates. Their outputs are concatenated together and are sent to the next Conv. layer.} 
    \label{details1} 
\end{figure*}

\subsection{L3C-Stereo Overview}
Fig. \ref{L3C-stereo} shows the massive proposed structure. The whole model focuses on predicting a distribution for each subpixel\footnote{A subpixel is the smallest unit of an image in a single channel. A pixel contains three subpixels for RGB images.}. Instead of directly predicting the distribution on $\boldsymbol{X}_{L}$ and $\boldsymbol{X}_{R}$, several auxiliary feature maps are used to simplify this task. Fig. \ref{details1} shows an encoder-decoder flow and the probability estimation module under scale $s$. Assuming that the $\boldsymbol{X}_{L}$ and $\boldsymbol{X}_{R}$ get a shape of $3\times H\times W$, after $S$ encoder modules we get auxiliary feature maps $\boldsymbol{Z}_{L}^{(s)}$ and $\boldsymbol{Z}_{R}^{(s)}$, where $s=1,2...S$, with a shape of $C\times H/2^{s} \times W/2^{s}$ ($C=5$ in all models appearing in this paper).

In this model, we jointly predict:
\begin{equation}
\begin{aligned}
&p(\boldsymbol{X}_{L},\boldsymbol{Z}_{L}^{(1)},...,\boldsymbol{Z}_{L}^{(S)})=\\
&\qquad p(\boldsymbol{X}_{L}\mid \boldsymbol{Z}_{L}^{(1)},...,\boldsymbol{Z}_{L}^{(S)})\prod_{s=1}^{S}p(\boldsymbol{Z}_{L}^{(s)}\mid \boldsymbol{Z}_{L}^{(s+1)},...,\boldsymbol{Z}_{L}^{(S)}),
\end{aligned}
\end{equation}

\begin{equation}
\begin{aligned}
&p(\boldsymbol{X}_{R},\boldsymbol{Z}_{R}^{(1)},...,\boldsymbol{Z}_{R}^{(S)}\mid \overline{\boldsymbol{X}}_{R},\overline{\boldsymbol{Z}}_{R}^{(1)},...,\overline{\boldsymbol{Z}}_{R}^{(S-1)})=\\
&\qquad p(\boldsymbol{X}_{R}\mid (\boldsymbol{Z}_{R}^{(1)},\overline{\boldsymbol{X}}_{R}),...,(\boldsymbol{Z}_{R}^{(S)},\overline{\boldsymbol{Z}}_{R}^{(S-1)}))\cdot \\
&\qquad \prod_{s=1}^{S-1}p(\boldsymbol{Z}_{R}^{(s)}\mid (\boldsymbol{Z}_{R}^{(s+1)},\overline{\boldsymbol{Z}}_{R}^{(s)}),...,(\boldsymbol{Z}_{R}^{(S)},\overline{\boldsymbol{Z}}_{R}^{(S-1)}))p(\boldsymbol{Z}_{R}^{(S)}).
\end{aligned}
\end{equation}
Where $\overline{\boldsymbol{X}}_{R}$ and $\overline{\boldsymbol{Z}}_{R}^{(s)}$ are the known from the $\texttt{Warp}$ modules, and $p(\boldsymbol{Z}_{L}^{(S)})$ and $p(\boldsymbol{Z}_{R}^{(S)})$ follow a \textbf{uniform distribution}.

\textbf{Algorithm pipeline}: Taking two view images as inputs, the L3C-Stereo method employs three encoder modules to extract the multi-scaled feature representation $\widetilde{\boldsymbol{Z}}^{(s)}$. A quantization module is used to discrete the outputs, so that the value of $\widetilde{\boldsymbol{Z}}^{(s)}$ can be transformed to $\boldsymbol{Z}^{(s)}$(which can take values of 0~255, just like an RGB image). Then, three decoder modules are used to generate feature maps $\boldsymbol{f}^{(s)}$ for the probability estimation. 
Until this point, the structure is very similar to the U-Net and Pyramid Network, except for the use of the quantization module which would lead to information loss. After that, for the left view, the process is relatively simple. Receiving $\boldsymbol{f}_{L}^{(s)}$, module $\texttt{P}_{L}^{(s)}$ predicts a conditional distribution $p(\boldsymbol{Z}_{L}^{(s-1)}\mid \boldsymbol{f}_{L}^{(s)})$. Since $\boldsymbol{f}_{L}^{(s)}$ contains the information from $\boldsymbol{Z}_{L}^{(s)}, \boldsymbol{Z}_{L}^{(s+1)},...,\boldsymbol{Z}_{L}^{(S)}$, this conditional distribution is equal to $p(\boldsymbol{Z}_{L}^{s-1}\mid \boldsymbol{Z}_{L}^{(s)}, \boldsymbol{Z}_{L}^{(s+1)},...,\boldsymbol{Z}_{L}^{(S)})$. For the right view, $\boldsymbol{f}_{L}^{(s)}$ and $\boldsymbol{f}_{R}^{(s)}$ are employed to generate the cost volume, which is then taken by the  $\texttt{Warp}^{(s)}$ module, along with the aggregated cost volume from $\texttt{Warp}^{(s+1)}$, to estimate the disparity map at this scale. Based on that, we warp $\boldsymbol{Z}_{L}^{(s-1)}$ into $\overline{\boldsymbol{Z}}_{R}^{(s-1)}$, expecting $\overline{\boldsymbol{Z}}_{R}^{(s-1)}$ and $\boldsymbol{Z}_{R}^{(s-1)}$ to be highly similar. Conducting Conv. $1\times1$ for the fusion between $\boldsymbol{f}_{R}^{(s)}$ and $\overline{\boldsymbol{Z}}_{R}^{(s-1)}$, we generate $\overline{\boldsymbol{f}}_{R}^{(s)}$, which is the fine-tuned version of $\boldsymbol{f}_{R}^{(s)}$. Finally, the same operation as with the left view is used to estimate the conditional probability distribution.

\begin{figure*}[ht] 
    \centering 
    \includegraphics[width=0.9\textwidth]{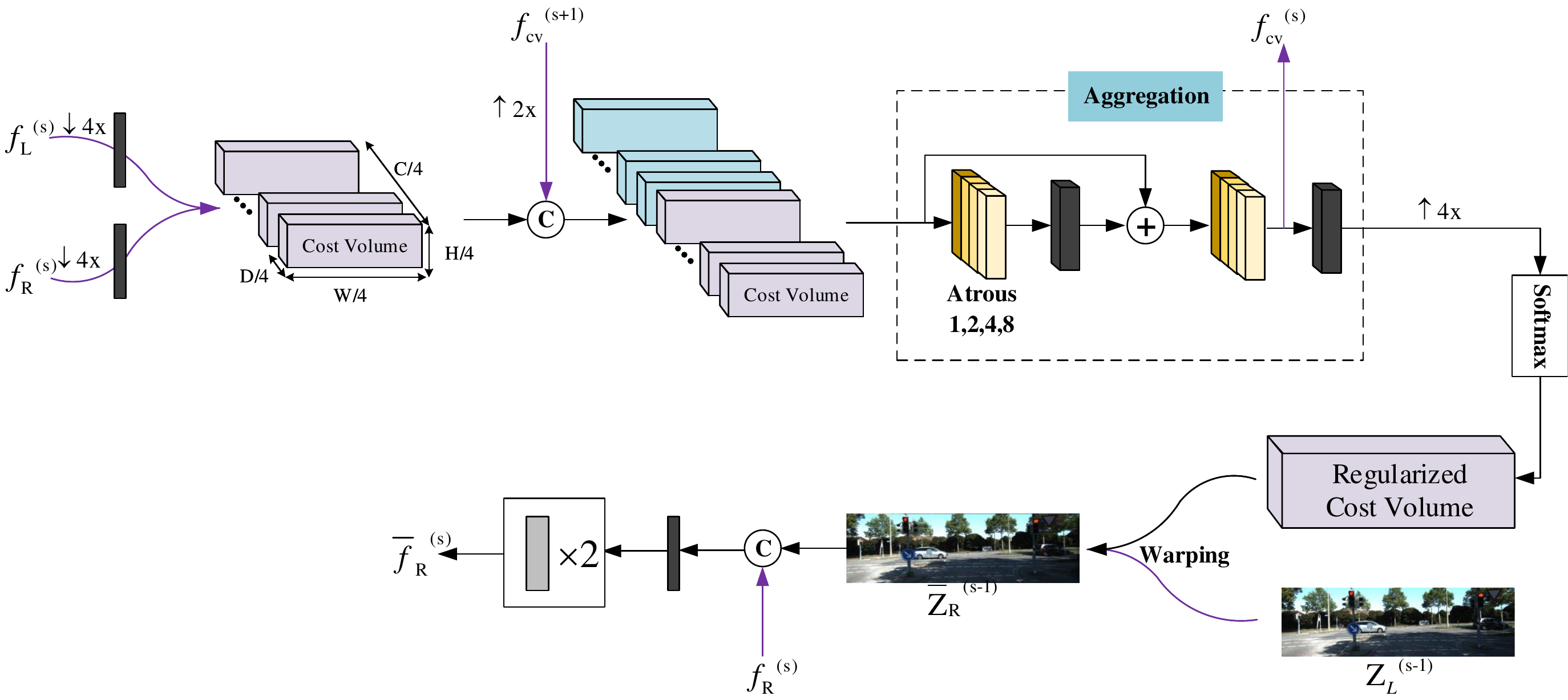} 
    \caption{Architecture details for $\texttt{Warp}^{(s)}$. $\downarrow/ \uparrow n\texttt{x}$ represents the interpolation with $n$ times under a bilinear or trilinear mode. The purple arrows represent the inputs for $\texttt{Warp}^{(s)}$ and the output for $\texttt{Warp}^{(s-1)}$. ``C'' refers to the concatenate operation. The slender cube represents a 3D Conv. layer, and the ``Aggregation'' structure is similar to $\texttt{P}^{(s)}$.} 
    \label{details2} 
\end{figure*}

\subsection{Quantization}
To discrete the value $\widetilde{z}\in \widetilde{\boldsymbol{Z}}^{(s)}$, we use the scalar quantization approach proposed in \cite{mentzer2018conditional}. Based on a list of quantized scalars $\mathbb{L}=\left \{ l_{1}, l_{2}, ..., l_{L} \right \}\subset \mathbb{R}$, the hard manner directly uses the nearest neighbor assignment algorithm:
\begin{equation}
\begin{aligned}
z=Q_{h}(\widetilde{z})=\mathbb{L}\left [ argmin_{j=1}^{L}\left | l_{j}-\widetilde{z} \right | \right ].
\end{aligned}
\end{equation}
However, this manner is non-differentiable. A soft manner solves this problem:
\begin{equation}
\begin{aligned}
z=Q_{s}(\widetilde{z})=\sum_{j=1}^{L}\frac{exp(-\sigma_{q}\left | l_{j}-\widetilde{z} \right | )}{\sum_{k=1}^{L}exp(-\sigma_{q}\left | l_{k}-\widetilde{z} \right | )}l_{j},
\end{aligned}
\end{equation}
where $\sigma_{q}$ is a softness hyper-parameter. During the training phase, we use $Q_{h}$ for the forward propagation and $Q_{s}$ for the backward propagation:
\begin{equation}
\begin{aligned}
z=\left \{ Q_{h}(\widetilde{z})-Q_{s}(\widetilde{z}) \right \}_{require\_no\_grad}+Q_{s}(\widetilde{z}).
\end{aligned}
\end{equation}
In this paper, we set $L=25$ and evenly distributed $\mathbb{L}$ in $[-1,1]$.

\subsection{Warp}
 A disparity map refers to the apparent pixel difference or motion between a pair of stereo images. As a result, we can move the left view pixel to the position where the right view pixel may appear based on the disparity in a pixel-wise manner. Then, the reconstructed right view can use the model to predict the distribution of the real right view pixel. To reduce the search space, the disparity is only estimated in the width dimension, and we use the sliding window on the width rather than on the full line data. 
 
 Specifically, assuming the disparity maximum, a.k.a. sliding widow, is $D$ pixels. The cost volume is generated by comparing the feature maps of both views with $D$ times of horizontal shift. Given $\boldsymbol{f}_{L}^{(s)}$ and $\boldsymbol{f}_{R}^{(s)}$ with a shape of $C\times H\times W$, the cost volume $CV$ can be written as (in a numpy form):
 \begin{equation}
\begin{aligned}
CV=\begin{bmatrix}
\left \| \boldsymbol{f}_{L}^{(s)}-\boldsymbol{f}_{R}^{(s)} \right \| &  &  & \\ 
\left \| \boldsymbol{f}_{L}^{(s)}[1:]-\boldsymbol{f}_{R}^{(s)}[:-1] \right \| & \boldsymbol{O} &  & \\ 
\left \| \boldsymbol{f}_{L}^{(s)}[2:]-\boldsymbol{f}_{R}^{(s)}[:-2] \right \| & \boldsymbol{O} & \boldsymbol{O} & \\ 
\vdots  & \vdots  & \vdots  & \vdots \\ 
\left \| \boldsymbol{f}_{L}^{(s)}[D-1:]-\boldsymbol{f}_{R}^{(s)}[:1-D] \right \| & \boldsymbol{O} & ... & \boldsymbol{O}
\end{bmatrix}.
\end{aligned}
\end{equation}
The $CV$ is a 4D tensor with a shape of $C \times D \times H \times W$. Given an exact pixel position, we will see a vector with $D$ numbers, and the smaller this number is, the more likely that the index of this number is the distance between the two views of the corresponding pixel. After that, an aggregation block is introduced for refining the cost volume among all channels $C$, generating a cost volume with a shape of $D \times H \times W$. Then, by conducting a softmax operation along $D$ dimension we will get a regularized cost volume, where a fiber along $D$ dimension represents the probability that this pixel in the right view matches all the pixels in the sliding window of the left view. Subsequently, the left view pixels are moved to the right view pixels in a soft manner: Given a pixel $\boldsymbol{z}_{R}$, a fiber $\boldsymbol{d}$ is extracted. Starting at this pixel, we then take D pixels $\boldsymbol{z}_{L}$ from the left view along the horizontal direction and apply the following equation:
\begin{equation}
\begin{aligned}
\overline{z}_{R}=sum(\boldsymbol{b}\cdot \boldsymbol{z}_{L}).
\end{aligned}
\end{equation}
In the end, the warped $\overline{\boldsymbol{Z}}_{R}^{(s)}$ is simply concatenated with $\boldsymbol{f}_{R}^{(s)}$ and the two are fused with several 2D conv. layers.

Considering the fact that the proposed model naturally generates multi-scaled feature maps, note that better results can be obtained by using the feature map of the previous scale to assist the module of the current scale in the prediction. Therefore, we introduce $\boldsymbol{f}_{cv}^{(s+1)}$ into $\texttt{Warp}^{(s)}$ before aggregation.

\subsection{Logistic Mixture Model}
In the prediction phase, a mixture of discrete logistics with $K$ components are used \cite{mentzer2019practical,salimans2017pixelcnn++} to model the probability mass function of each subpixel in $\boldsymbol{X}$ and $\boldsymbol{Z}^{(s)},s=1,2...,S$. Since the left and right views are handled in the same way, here the L and R subscripts will be omitted. 

A single logistic distribution is given by:
\begin{equation}
\begin{aligned}
p_l(z|\mu ,\sigma )=\frac{e^{-(z-\mu)/\sigma}}{\sigma(1+e^{-(z-\mu)/\sigma})^2}.
\end{aligned}
\end{equation}
For the discrete form, we take advantage of its cumulative distribution function to compute the cumulative probability of the discrete interval:
\begin{equation}
\begin{aligned}
p_l(z|\mu ,\sigma )=sigmoid(\frac{z+b/2-\mu}{\sigma })-sigmoid(\frac{z-b/2-\mu}{\sigma }),
\end{aligned}
\end{equation}
where $b$ is the discrete interval, $b=1$ for $X$ and $b=1/12$ for $Z^{(s)}$. 

Given a pixel $\boldsymbol{x}$ from $\boldsymbol{X}$, where $\left ( x_{i},i=1,2,3 \right )\in \boldsymbol{X}$, it is modeled in an auto-regression form:
\begin{equation}
\begin{aligned}
p(\boldsymbol{x}|\boldsymbol{f}^{(1)})=&p(x_1,x_2,x_3|\boldsymbol{f}^{(1)})\\
=&p_m(x_1|\boldsymbol{f}^{(1)})\cdot p_m(x_2|\boldsymbol{f}^{(1)},x_1)\cdot p_m(x_3|\boldsymbol{f}^{(1)},x_1,x_2).
\end{aligned}
\end{equation}
Where $p_m$ is a mixture of logistic distributions $p_l$. $p_m$ contains $K$ components, and the $k$th component has three parameters: weight $\pi _i^k$, mean $\mu _i^k$ and variance $\sigma _i^k$. Then, we get:
\begin{equation}
\begin{aligned}
p_m(x_1|\boldsymbol{f}^{(1)})=\sum _k \pi _1^k p_l(x_1|\mu _1^k,\sigma _1^k)\\
p_m(x_2|\boldsymbol{f}^{(1)},x_1)=\sum _k \pi _2^k p_l(x_2|\widetilde{\mu} _2^k,\sigma _2^k)\\
p_m(x_3|\boldsymbol{f}^{(1)},x_1,x_2)=\sum _k \pi _3^k p_l(x_3|\widetilde{\mu} _3^k,\sigma _3^k).
\end{aligned}
\end{equation}
As previously mentioned, on the topic of auto-regression, the model provides the coefficients $\lambda _\alpha ^k,\lambda _\beta  ^k,\lambda _\gamma  ^k$ for a better prediction of $\mu$:
\begin{equation}
\begin{aligned}
\widetilde{\mu} _2^k&=\mu _2^k+\lambda _\alpha ^k x_1\\
\widetilde{\mu} _3^k&=\mu _2^k+\lambda _\beta ^k x_1 + \lambda _\gamma  ^k x_2.
\end{aligned}
\end{equation}
Given a pixel $\boldsymbol{z}^{(s)}$ from $\boldsymbol{Z}^{(s)}$ under scale $s$, where $\left ( z_{i}^{(s)},i=1,2,3,4,5 \right )\in \boldsymbol{Z}^{(s)}$, we get
\begin{equation}
\begin{aligned}
p(\boldsymbol{z}^{(s)}|\boldsymbol{f}^{(s+1)})&=p(z_1^{(s)},z_2^{(s)},z_3^{(s)},z_4^{(s)},z_5^{(s)}|\boldsymbol{f}^{(s+1)})\\
&=\prod_{i=1}^{5}p_m(z_i^{(s)}|\boldsymbol{f}^{(s+1)})
\end{aligned}
\end{equation}
and
\begin{equation}
\begin{aligned}
p_m(z_i^{(s)}|\boldsymbol{f}^{(s+1)})=\sum _k \pi _{i}^{k(s)} p_l(z_i^{(s)}|\mu _i^{k(s)},\sigma _i^{k(s)}).
\end{aligned}
\end{equation}
We do not auto-regress $\mu$ when predicting the distribution of $Z^{(s)}$. 

All of $\mu$, $\pi$, $\sigma$ and $\lambda$ (if needed) are obtained using the last 2D Conv. $1\times 1$ layer of module $\texttt{P}$. Then, by carrying out the above operation, we will get the probability tensor with a shape of $3 \times H \times W \times 256$ for RGB images and $5 \times H^{(s)} \times W^{(s)} \times L$ for $\boldsymbol{Z}^{(s)}$.

\subsection{Loss}
Equation \ref{cross-entropy} describes the difference between the two distributions. Because the model gives the probability of every value of each subpixel, the subpixel value can be changed to a one-hot form as the target distribution. Then, the loss function is obtained as follows: 
\begin{equation}
\begin{aligned}
\mathcal{L}&=H(\widetilde{p}(\boldsymbol{X}_{L}),p(\boldsymbol{X}_{L}\mid \boldsymbol{Z}_{L}^{(1)},...,\boldsymbol{Z}_{L}^{(S)}))+\\\\
&\sum_{s=1}^{S-1}H(\widetilde{p}(\boldsymbol{Z}_{L}^{(s)}),p(\boldsymbol{Z}_{L}^{(s)}\mid \boldsymbol{Z}_{L}^{(s+1)},...,\boldsymbol{Z}_{L}^{(S)}))+\\
&H(\widetilde{p}(\boldsymbol{X}_{R}),p(\boldsymbol{X}_{R}\mid (\boldsymbol{Z}_{R}^{(1)},\overline{\boldsymbol{X}}_{R}),...,(\boldsymbol{Z}_{R}^{(S)},\overline{\boldsymbol{Z}}_{R}^{(S-1)})))+\\
&\sum_{s=1}^{S-1}H(\widetilde{p}(\boldsymbol{Z}_{R}^{(s)}),p(\boldsymbol{Z}_{R}^{(s)}\mid (\boldsymbol{Z}_{R}^{(s+1)},\overline{\boldsymbol{Z}}_{R}^{(s)}),...,(\boldsymbol{Z}_{R}^{(S)},\overline{\boldsymbol{Z}}_{R}^{(S-1)})))\\
&=-(log_2(p(\boldsymbol{X}_{L}\mid \boldsymbol{Z}_{L}^{(1)},...,\boldsymbol{Z}_{L}^{(S)}))+\\
&\sum_{s=1}^{S-1}log_2(p(\boldsymbol{Z}_{L}^{(s)}\mid \boldsymbol{Z}_{L}^{(s+1)},...,\boldsymbol{Z}_{L}^{(S)}))+\\
&log_2(p(\boldsymbol{X}_{R}\mid (\boldsymbol{Z}_{R}^{(1)},\overline{\boldsymbol{X}}_{R}),...,(\boldsymbol{Z}_{R}^{(S)},\overline{\boldsymbol{Z}}_{R}^{(S-1)})))+\\
&\sum_{s=1}^{S-1}log_2(p(\boldsymbol{Z}_{R}^{(s)}\mid (\boldsymbol{Z}_{R}^{(s+1)},\overline{\boldsymbol{Z}}_{R}^{(s)}),...,(\boldsymbol{Z}_{R}^{(S)},\overline{\boldsymbol{Z}}_{R}^{(S-1)})))).
\end{aligned}
\end{equation}

\begin{table*}[ht]
\caption{Comparison. ``bpsp'' means bits per subpixel. The bitcost without compression is 8 bits.} 
\label{compare} 
\centering 
\begin{tabular}{|c|c|c|c|c|c|c|c|c|c|}
\hline
\multirow{2}{*}{bpsp} & \multicolumn{3}{c|}{KITTI2012} & \multicolumn{3}{c|}{KITTI2015} & \multicolumn{3}{c|}{Flythings3D} \\ \cline{2-10} 
 & Left view & Right view & All view & Left view & Right view & All view & Left view & Right view & All view \\ \hline
PNG & 4.639 & 4.617 & 4.628 & 4.706 & 4.484 & 4.595 & 3.721 & 3.721 & 3.721 \\ \hline
WebP & 4.156 & 4.018 & 4.087 & 4.062 & 3.797 & 3.929 & 2.735 & 2.736 & 2.736 \\ \hline
FLIF & 4.026 & 3.884 & 3.955 & 4.077 & 3.806 & 3.941 & 2.601 & 2.602 & 2.602 \\ \hline
L3C & 3.946 & 3.803 & 3.875 & 3.611 & 3.339 & 3.475 & 2.587 & 2.589 & 2.588 \\ \hline
L3C-Stereo & \textbf{3.946} & \textbf{3.727} & \textbf{3.837} & \textbf{3.611} & \textbf{3.237} & \textbf{3.424} & \textbf{2.587} & \textbf{2.297} & \textbf{2.442} \\ \hline
\end{tabular}
\end{table*}

\section{Experiments}
In this section, we will aim to demonstrate that the L3C-Stereo technique outperforms all the compared methods, including PNG, WebP, FLIF and L3C. Ablation studies on  supervised/self-supervised disparity map estimation, on the maximum disparity and on the stereo matching structure are conducted to explore the influence of network hyper-parameters on the method's performance.

\subsection{Experiment Setup}
\textbf{Datasets} The methods are evaluated using three datasets:
\begin{enumerate}
\item KITTI2012: A real-world dataset with images collected using two high-resolution colour video cameras from a standard station wagon. It consists of 388 training image pairs and 390 testing image pairs, each one of size $376 \times 1240$.
\item KITTI2015\cite{7298925}: A real-world dataset with images collected using two high-resolution colour video cameras from a standard station wagon. It consists of 400 training image pairs and 400 testing image pairs, each one of size $376 \times 1240$.
\item Scene Flow\cite{mayer2016large}: A large synthetic dataset. One of its subsets, FlyThings3D, contains 22,390 training image pairs and 4,370 testing image pairs, each one of size $540 \times 960$. This subset was picked to be used as our dataset rather than the full Scene Flow dataset, since it was the only one out of the subsets in Scene Flow containing a testing set. 

\end{enumerate}
\textbf{Training} The proposed model was implemented with Pytorch\cite{paszke2017automatic}. All models were optimized with RMSProp\cite{tieleman2017divide}, with a batch size of 2. The images were randomly cropped to $128 \times 512$. The initial learning rate was 1e-4 and decayed by a factor of 0.75 every 400 epochs for the KITTI2012/2015 datasets and every 10 epochs for the Scene Flow dataset. For a fair comparison with L3C, we firstly trained L3C following the same network setup used in \cite{mentzer2019practical}, and the learning rate decayed by a factor of 0.75 every 1500 epochs for the KITTI2012/2015 datasets and every 30 epochs for the Scene Flow dataset. After L3C converged completely, the weights $\texttt{E}^{(s)}$, $\texttt{D}^{(s)}$ and $\texttt{P}_L^{(s)}$($s=1,2,3$) were replaced into some modules in L3C-Stereo and were frozen. The default maximum disparity was 64. All models is trained and tested on NVIDIA RTX 3090 unless otherwise specified.

\subsection{Comparison} \label{Comparison}

The compared methods include PNG, WebP, FLIF and L3C. Results are shown in Table \ref{compare}. The evaluation criterion used is bit per sub-pixel (bpsp). Since part of the weights used in the training of L3C-Stereo came from the trained L3C, the evaluation results of the L3C and L3C-Stereo left view were the same. For the right view, the proposed method outperforms all other compared methods, including L3C, where L3C-Stereo outperforms it by a small margin of $2.25\%$ on KITTI2012 and $3.15\%$ on KITTI2015, and by a larger margin of $12.7\%$ on the Scene Flow dataset. 

\begin{figure*}[ht] 
    \centering 
    \includegraphics[width=0.9\textwidth]{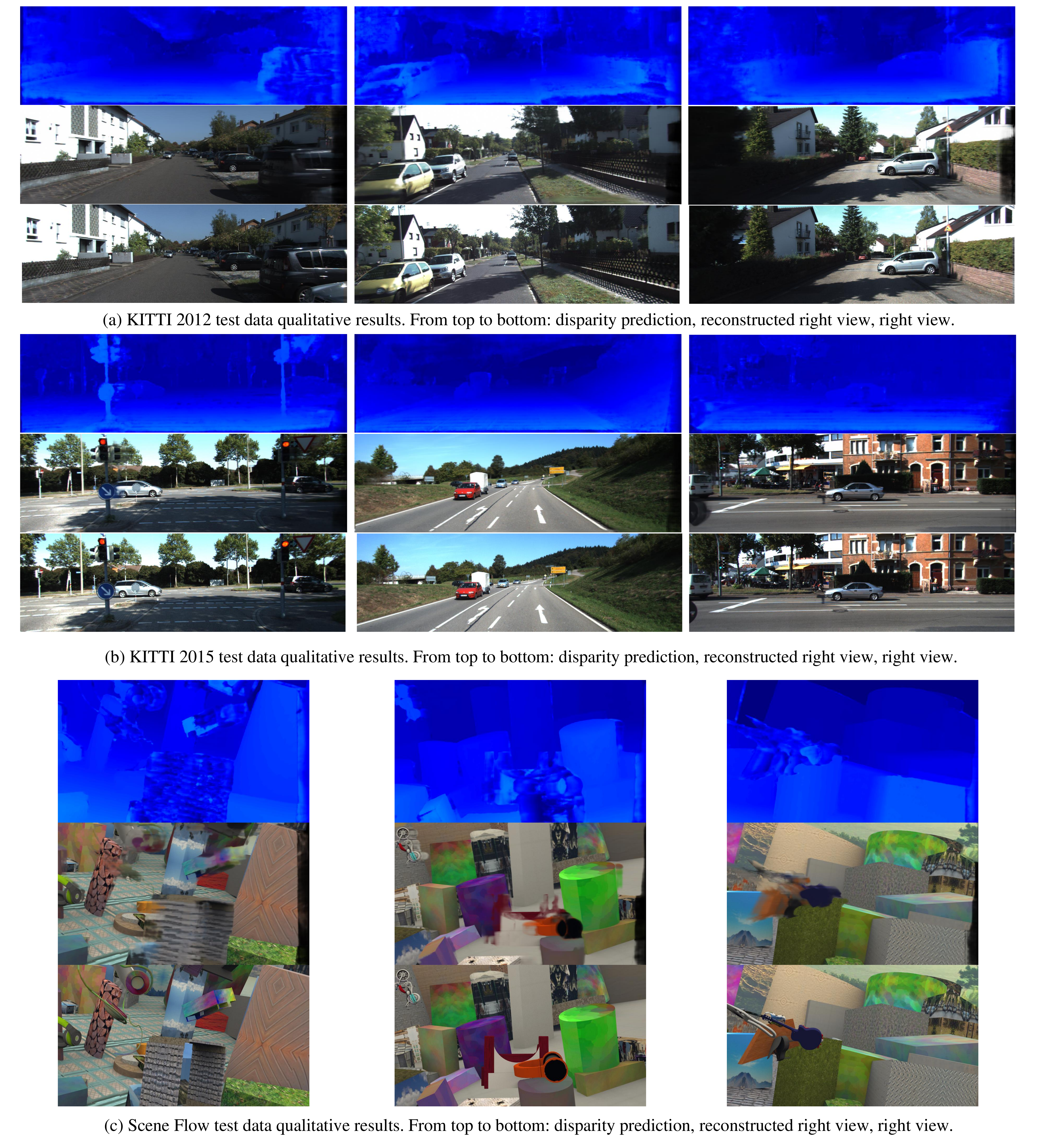} 
    \caption{Qualitative results from L3C-Stereo.} 
    \label{Qualitative} 
\end{figure*}

Fig. \ref{Qualitative} gives several qualitative results with regards to the disparity prediction and the reconstructed right view. The samples were randomly chosen from the testing set. Table \ref{quantitative} gives the image quality of the reconstructed right view (PSNR, SSIM) and the quality of the disparity map ($>$3px, EPE).  ``$>$3px'' represents the percentage of pixels for which the disparity error is more than three pixels (or equivalently, more than $5\%$) from all testing images. End-point error (EPE) is the mean absolute disparity error in pixels. From the results, it can be seen that the reconstructed right view from the real-world image has a higher PSNR and SSIM than the one from the synthetic image, but nevertheless the compression effect is not significantly improved. We think that real-world images have gentler pixel features, resulting in a gentler distribution of pixels predicted by the network. Since the KITTI datasets do not provide a right view disparity map, the quality of the disparity prediction cannot be calculated. For the disparity error, the results on the Scene Flow dataset are unexpected. More than $30\%$ of the image area has an error of more than 3 pixels, as can be seen from the relatively blurry reconstruction of nearby objects in Fig. \ref{Qualitative}.(c). The disparity of these objects exceeds our default 64 pixels, which leads to the above result. After using a larger maximum disparity, this phenomenon is alleviated.

\begin{table}[ht]
\caption{quantitative results from L3C-Stereo.} 
\label{quantitative}
\centering 
\begin{tabular}{|c|c|c|c|c|}
\hline
 & PSNR & SSIM & \textgreater{}3px(\%) & EPE \\ \hline
KITTI 2012 & 21.23 & 0.9225 & - & - \\ \hline
KITTI 2015 & 22.73 & 0.9531 & - & - \\ \hline
Scene Flow & 19.01 & 0.7231 & 31.80 & 21.08 \\ \hline
\end{tabular}
\end{table}

\subsection{Ablation study}
In this experiment, the impact of the additional hyper-parameters on L3C-Stereo was evaluated on the Scene Flow dataset, since it is the only dataset which could provide a right view disparity map. All training settings remained the same as those in Section \ref{Comparison}.

For the disparity estimation, current studies are based on two approaches to train the networks; one involves using a ground-truth disparity map for supervised training, while the other relies on minimising the warping error of the reconstructed view. L3C-Stereo adopts the second method, which doesn't require the use of any ground-truth disparity map. To discuss whether the supervised method can obtain a better performance, a term is added to the loss function:
\begin{equation}
\begin{aligned}
Loss=&\mathcal{L}+ \lambda ( \frac{\alpha _0}{N}\sum_{n=1}^{N}\left \| d_n- \widehat{d}_n^{0} \right \|_1+\\
&\frac{\alpha _1}{N}\sum_{n=1}^{N}\left \| d_n- \widehat{d}_n^{1} \right \|_1 +\frac{\alpha _2}{N}\sum_{n=1}^{N}\left \| d_n- \widehat{d}_n^{2} \right \|_1 )
\end{aligned}
\end{equation}
where $N$ is the number of pixels, $\lambda$ is a trade-off parameter and we set $\lambda=0.01$, $\alpha _0=1.0$, $\alpha _1=0.5$, $\alpha _2=0.25$. $d_n$ is the ground-truth disparity, $\widehat{d}_n^{s} (s=0,1,2)$ is the predicted disparity under three scale. Before calculating, the low-scale predicted disparity map is up-sampled in bilinear mode so as to be of the same size as the ground-truth disparity map.

Other three ablation settings include:
\begin{enumerate}
\item Disabling the warping function and replacing the reconstructed right view image with the left view image.
\item Disabling $\boldsymbol{f}_{cv}^{(s)}$ from module $\texttt{Warp}^{(s)}$ and doubling the output channels to be consistent with the previous layer.
\item The options for maximum disparity include 64, 128 and 192.
\end{enumerate}

\begin{table*}[ht]
\caption{Ablation Study on L3C-Stereo.} 
\label{Ablation study}
\centering 
\begin{tabular}{|c|c|c|c|c|c|c|c|}
\hline
Disparity estimation & Ablation & D\_max & bpsp & PSRN & SSIM & \textgreater{}3px(\%) & EPE \\ \hline
- & Baseline & - & 2.589 & - & - & - & - \\ \hline
- & Use left view & - & 2.539 & - & - & - & - \\ \hline
\multirow{4}{*}{Self-supervised} & Disable $\boldsymbol{f}_{cv}$ & 64 & 2.306 & 19.53 & 0.7413 & 32.10 & 21.08 \\ \cline{2-8} 
 & \multirow{3}{*}{Default} & 64 & 2.296 & 19.01 & 0.7231 & 31.80 & 21.08 \\ \cline{3-8} 
 &  & 128 & 2.231 & \textbf{21.96} & \textbf{0.9110} & 19.41 & 12.94 \\ \cline{3-8} 
 &  & 192 & \textbf{2.226} & 21.39 & 0.8959 & 17.79 & 13.19 \\ \hline
\multirow{3}{*}{Supervised} & \multirow{3}{*}{Default} & 64 & 2.310 & 18.36 & 0.7163 & 24.56 & 19.76 \\ \cline{3-8} 
 &  & 128 & 2.251 & 19.89 & 0.8623 & 9.89 & 9.73 \\ \cline{3-8} 
 &  & 196 & 2.236 & 19.83 & 0.8663 & \textbf{7.95} & \textbf{7.95} \\ \hline
\end{tabular}
\end{table*}

\begin{figure*}[ht] 
    \centering 
    \includegraphics[width=0.9\textwidth]{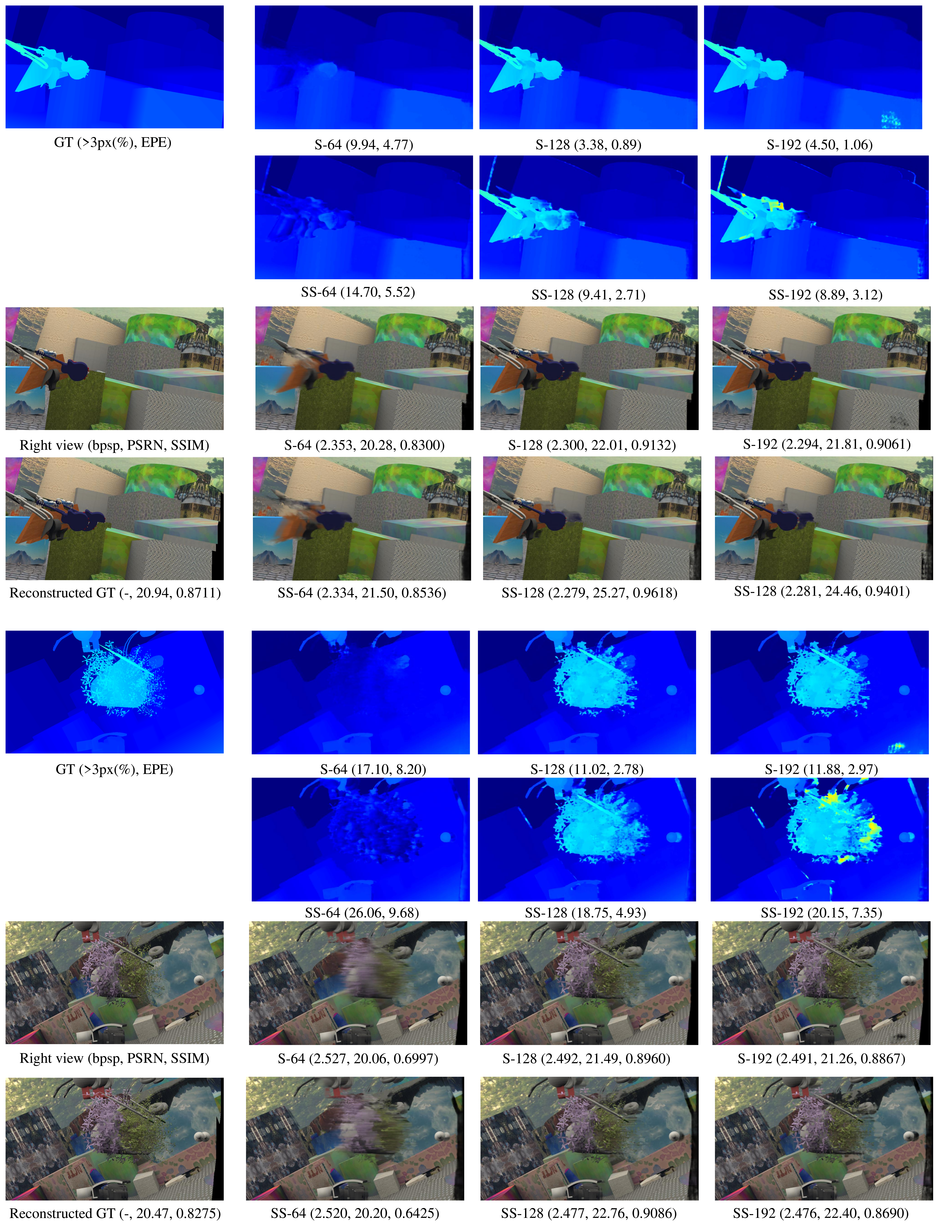} 
    \caption{Case study. Two cases from the testing set are shown, in which one is easily compressed, and the other is difficult to compress. ``S'' means ``Supervised'' and ``SS'' means ``Self-supervised''. The $n$ in ``SS-$n$'' and ``S-$n$'' refers to the maximum disparity. ``GT'' means ground-truth. The numbers in the brackets represent the content in brackets under the left-hand side image of the same kind. The reconstructed GT is the result of warping the left view based on the GT disparity map.} 
    \label{case study} 
\end{figure*}

All the results are presented in Table \ref{Ablation study}. ``Baseline'' refers to the L3C result, using none of the disparity or left view information. ``Use left view'' corresponds to point 1 above. ``Disable $\boldsymbol{f}_{cv}$'' corresponds to point 2 above. ``D\_max'' refers to maximum disparity. ``bpsp'' means bits per sub-pixel for the right view images. ``PSNR'' and ``SSIM'' evaluate the quality of the reconstructed right view. Finally, ``$>$3px(\%)'' and ``EPE'' evaluate the quality of the disparity map. Fig. \ref{case study} shows the disparity map and the reconstructed right view from the two cases, in which one is easy to compress and the other is hard to compress. In the figure, ``S'' means ``Supervised'' and ``SS'' means ``Self-supervised''. The reconstructed ground-truth (GT) is the result obtained by warping the left view using the ground-truth disparity map, where you may notice some ghosting in the image. ``bpsp'' is the compressed right view result based on the shown reconstructed right view. The brighter the disparity map, the greater the disparity and therefore the closer the object is.

\begin{table}[]
\caption{Comparison between the different disparity estimation methods.} 
\label{stereo matching}
\centering 
\begin{tabular}{|c|c|c|}
\hline
 & $>$3px(\%) & EPE \\ \hline
L3C-Stereo & 7.95 & 7.95 \\ \hline
GC-Net\cite{kendall2017end} & 9.34 & 2.51 \\ \hline
PSMNet\cite{chang2018pyramid} & 3.80 & 1.09 \\ \hline
GwcNet\cite{guo2019group} & 3.30 & 0.765 \\ \hline
\end{tabular}
\end{table}

Results show that all the L3C-Stereo structures outperform, by a large margin, those which disable the left view or disparity information. Compared with the baseline, directly using a left view image only leads to a small increase in performance. Since there is a pixel level difference between the two views, although view overlap can occur, it still provides a negative gain for the right view probability estimation. Disabling $\boldsymbol{f}_{cv}^{(s)}$ also slightly degrades the performance.

As can be observed both in Table \ref{Ablation study} and in Fig. \ref{case study}, all models trained through self-supervision widely outperform the ones trained through supervision. For both self-supervised and supervised models, a higher maximum disparity significantly results in a more accurate disparity map, an improved reconstructed right view and a lower bpsp. However, blindly increasing the maximum disparity also has a boundary effect, which may even lead to a worse model.

The common settings for the disparity estimation task are 192 maximum disparity and supervision. For a fair comparison, some recent disparity estimation results are compared with our ``S-192'' L3C-Stereo the Table \ref{stereo matching}. Our proposed method, as expected, does not perform very well. Aiming at a better compression, the encoder modules in L3C-Stereo prefer to extract feature maps that are easier to compress. Furthermore, the feature maps used for disparity estimation are all derived from the auxiliary feature maps that have been squeezed along the channel, which makes it harder for the model to predict an accurate disparity map. However, our model can still estimate a good disparity map, which we think is partly due to the use of a multi-scale structure similar to PSMNet.

One may notice that self-supervised disparity maps are worse than supervised ones, which is reasonable, since supervised methods require ground-truth, which contributes to the strong effect of supervision. On the other hand, the reconstructed right view under self-supervision is better than the one under supervision. We think that there are some occlusions in the left view, as it was impossible to reconstruct these areas using the information from the disparity map and the left view. Expecting a better reconstruction quality, self-supervision is not attempting to estimate the best disparity map; instead, it searches for the position in the window with the maximum disparity length that is most likely to be the pixel value of the right view. In Fig. \ref{case study}, compared with the right view, there is significant ghosting in the reconstructed GT. In fact, these ghostly objects on the right side appear in the left view image. Since these parts of the area are occluded, they cannot be reconstructed. As a consequence, the reconstructed GT shows a low PSNR and SSIM. For instance, in case two, SS (self-supervision) reconstructed right view results show a shadow near the ball, and the colour of the shadow is brighter here. Due to occlusion, the left shadow is the information related to reconstructing the ball in the right view, and the right shadow is the information related to reconstructing the occluded area. In the reconstructed right view of ``SS-128'' and ``SS-192'', it can be seen that the results are relatively better than ``S-$n$'' models in the ball area.

\subsection{Complexity}
Table \ref{Complexity} gives the model parameters, FLOPs and runtime for a pair of $512 \times 512$ images (CPU: Intel i5 6500, GPU: Nvidia GTX 980Ti). Compared with L3C, our method does not significantly increase the computational or time complexities. For the different maximum disparity, it does not bring in a lot of FLOPs, and instead brings a greater time consumption. We think that the warping module does not use the GPU continuously; some operators, such as generating cost volume and warping images, use a lot of CPU, which could be optimized.

\begin{table*}[h]
\caption{Complexity Comparison.} 
\label{Complexity}
\centering 
\begin{tabular}{|c|c|c|c|c|c|c|}
\hline
\multirow{2}{*}{Model} & \multicolumn{3}{c|}{Encoding} & \multicolumn{3}{c|}{Decoding} \\ \cline{2-7} 
 & Param(M) & FLOPs(G) & Runtime(s) & Param(M) & FLOPs(G) & Runtime(s) \\ \hline
L3C & 4.933 & 352.275 & 0.725 & 2.741 & 226.594 & 0.665 \\ \hline
L3C-Stereo(64) & \multirow{3}{*}{5.935} & 421.100 & 0.850 & \multirow{3}{*}{3.743} & 295.419 & 0.788 \\ \cline{1-1} \cline{3-4} \cline{6-7} 
L3C-Stereo(128) &  & 437.548 & 0.937 &  & 311.868 & 0.871 \\ \cline{1-1} \cline{3-4} \cline{6-7} 
L3C-Stereo(192) &  & 453.997 & 1.026 &  & 328.316 & 0.957 \\ \hline
\end{tabular}
\end{table*}

\section{Conclusion}
In this paper, we propose the L3C-Stereo method for the compression of stereo images. By leveraging the disparity map, our method outperforms the compared methods on three datasets. In the experiments, we show that a significant margin can be achieved by using the reconstructed right view instead of the left view. Furthermore, the increase of the maximum disparity hyper-parameter will lead to a decrease in the running speed of the model and a better compression effect. However, blindly increasing this parameter will also lead to a decrease in the model performance. What's more, the by-products of the model, the disparity map, still is of a relatively good quality. Future work should investigate the reconstruction under occlusion, which should use the texture features near occlusion, and a more efficient way to warp the left view.

\section{Acknowledgement}
This research was supported by the National Key R\&D Program of China (No. 2017YFE0129700), the National Natural Science Foundation of China (Key Program) (No. 11932013), the National Natural Science Foundation of China (No. 62006045), the Tianjin Natural Science Foundation for Distinguished Young Scholars (No. 18JCJQJC46100), Japan Society for the Promotion of Science ( Grant-in-Aid for Early-Career Scientists 20K19875).

\ifCLASSOPTIONcaptionsoff
  \newpage
\fi



%

\bibliographystyle{ieeetr}
\bibliography{ref.bib}
%








\end{document}